\documentclass{pasj00}
\begin{document}
\SetRunningHead{S. Sasaki and N.Y. Yamasaki}{Scaling Relation for
Clusters of Galaxies}
\Received{2001/7/31}%{yyyy/mm/dd}
\Accepted{2001/11/5}%{yyyy/mm/dd}

\title{Scaling Relation to Understand Non-Detection of Cold Gas at
the Cluster Center}

%%% begin:list of authors
\author{%
Shin \textsc{Sasaki} and Noriko Y. \textsc{Yamasaki}}
\affil{Department of Physics, Tokyo Metropolitan University,
Hachioji, Tokyo 192-0397}
\email{sasaki@phys.metro-u.ac.jp}
\email{yamasaki@phys.metro-u.ac.jp}
%%% end:list of authors

%% `\KeyWords{}' always has to be placed before `\maketitle'.
 %Do NOT move this preamble from here!

\maketitle

\begin{abstract}
 Recent XMM-Newton observations of clusters of galaxies have
 indicated the soft X-ray spectra to be inconsistent with the simple
 isobaric  cooling flow model. There is almost no feature of the
 cold gas expected from the model.  This shows that we have not yet
 understood the physics of the hot gas in clusters of galaxies
 well. 
 A quantitative evaluation of the behavior of gas cooling is important
 not only for understanding the clusters, themselves, but also for
 studying cosmology and galaxy formation. To clarify the problem of
 this reported discrepancy, we have studied scaling relations for
 clusters of  galaxies based on the self-similarity assumption.
 We also propose an observational strategy to solve this problem.
\end{abstract}

\section{Introduction}

Previous soft X-ray imaging studies with medium spectral resolution
have shown evidence of cooling flows in many clusters of galaxies.
In many cases, a simple isobaric cooling flow model
(\cite{jfet92}) was used, and showed a reasonable fit to the observed
results. 
However, recent XMM-Newton observations with high spectral resolution
(\cite{tam01}; \cite{kaa01}; \cite{pet01}; \cite{mbfb01}) showed that
the soft X-ray spectra is inconsistent with the simple isobaric
cooling flow model.  
In A1795, Tamura et al. (2001) showed that the XMM-Newton RGS spectrum
obtained from the cluster center can be described by an isothermal model
with $kT\sim$ 4 keV, and that the upper limit of the emission measure
of the cool component ($kT < 1$ keV) is only a few \%.
Kaastra et al. (2001) showed a significant lack of cool gas below
1.5 keV in the S\'{e}rsic 159-03 cluster of galaxies, where a cooling
flow of 230 $M_{\solar}$ yr$^{-1}$ has been indicated from the ROSAT
PSPC observation (\cite{all97}).  
Peterson et al. (2001) showed that the RGS energy spectrum at the
cluster center of A1835 requires a cut-off in the emission measure
distribution at 2.7 keV to obtain a good fit with the isobaric
cooling flow model.
In the Virgo cluster, Matsushita et al. (2001) reported that the
emission measure of the cool component below 1 keV is much lower
than the expected value from the cooling flow model.
In summary, although the central excess of the surface brightness
profile has suggested cooling flow and the existence of cold gas,
RGS high-resolution spectra at the center of cooling flow clusters
did not exhibit the expected emission lines from the cold gas,
and the temperature distribution had to be cut-off (cut-off
temperature, $T_{\rm cut}$) at 1--3 keV (hereafter, we call this
feature the ``lack of cold gas'' problem).
This indicates that we have not yet understood the physics of the hot
gas in clusters well.

A proper understanding of the process of gas cooling is important
in many respects.  First, it is necessary to discuss the cluster gas
evolution and its status (for example, whether the hot gas is in
hydrostatic equilibrium or not).
Second, it is a key to understand the dark matter properties and
evolution, since we cannot observe dark matter directly, and its
information is obtained only through the observed hot gas properties.
Third, clusters are important to study
cosmology (for example, to estimate the cosmological parameters), and
have been widely used in research.
In order to use clusters as a cosmological probe, we need to understand
the cluster properties well.
Fourth, a study of the cooling gas in clusters gives
important information for understanding galaxy formation and evolution,
because we can regard galaxies as small mass clusters as a first
approximation when we discuss galaxy formation and evolution. 
Then, cooling is more important in galaxies than in clusters,
since the cooling time of the hot gas in galaxy-size halos is shorter
than the cosmic time (this feature is different from the clusters'
case) and stars are formed from the cooled gas (we have obtained
various information about galaxies through stars).

Thus, it is important to understand the cluster hot gas properties.
In particular, we need to solve this ``lack of cold gas''
problem. There are several possibilities to solve this problem:
heating, thermal conduction, inhomogeneous metallicity, absorption by
cold gas, and non-standard cooling function, and so on (\cite{fab01};
\cite{pet01}).  These mechanisms have both advantages and
disadvantages, and it is difficult to judge which is the main
mechanism at present. 

It seems to be difficult to solve the ``lack of cold gas'' problem by
studying individual clusters, as in previous literature.  Thus, in this
Letter, we discuss this problem while considering correlations
between cluster properties.

\section{Scaling Relation}

In this section, we consider correlations between cluster properties,
particularly between $L_{\rm cool}$ (as defined below) and the gas
temperature (or cluster mass). 
First, we review the self-similarity model briefly and then
modify it in order to explain the observed luminosity--temperature
relation.
We then consider the cooling effect on the cluster properties. 

At first, we consider the whole properties of clusters. If shock
heating caused by gravitational collapse is the dominant
mechanism, it is thus naturally expected that clusters follow a
simple scaling model (\cite{kai86}).  We assume that clusters of
galaxies are spherically symmetric, and that the hot gas is isothermal and
in a steady state with hydrostatic equilibrium. We also assume that
the gas number density profile $n(r)$ is described by the
conventional $\beta$ model, $n(r) = n_0/ [1+(r/r_{\rm c})^2]^{3
\beta/2}$, with $\beta = 2/3$, and that the core radius $r_{\rm c}$ is
proportional to the cluster radius $R_{\rm cl}$: $r_{\rm c} \propto
R_{\rm cl}$. 
Assuming that the average mass density in clusters and the gas fraction
are identical in all clusters, we obtain the following relations:
\begin{equation}
 n_0 \propto M_{\rm cl}^0 ~(={\rm constant}), ~~~
 r_{\rm c} \propto M_{\rm cl}^{1/3}, ~~~
 T_{\rm vir} \propto M_{\rm cl}^{2/3},
\end{equation}
where $M_{\rm cl}$ and $T_{\rm vir}$ are the cluster mass and virial
temperature, respectively. Hereafter, we assume that the gas
temperature, $T_{\rm gas}$, equals the virial temperature ($T_{\rm gas}
\equiv T_{\rm vir}$) when gas cooling is neglected. The above relations
imply that $r_{\rm c} \propto R_{\rm cl} \propto T_{\rm gas}^{1/2}$.
Although whether the observational data satisfies this relation is
still in controversy; for example, Vikhlinin et al. (1999) found that
the radius of a fixed mean gas overdensity of 1000 is scaled to $T_{\rm
gas}^{1/2}$. 

If bremsstrahlung emission is the dominant emission
mechanism, the bolometric luminosity, $L_{\rm bol}$, is proportional to
the gas temperature squared: $L_{\rm bol} \propto {n_0}^2 {r_{\rm c}}^3
T_{\rm gas}^{1/2} \propto T_{\rm gas}^2$. However, the observed
luminosity--temperature relation is steeper than this relation; the
luminosity is roughly proportional to the gas temperature cubed:
$L_{\rm bol} \propto T_{\rm gas}^3$ (e.g., \cite{dsj93}).
There are many arguments to explain this discrepancy.
We can divide them into roughly two groups.
One is that the gas fraction varies with the cluster mass;
the other is that 
the gas profile deviates from the self-similarity [equation (1)].
For example, 
the core radius is not proportional to $R_{\rm cl}$, or the cluster
radius is not proportional to $T_{\rm gas}^{1/2}$.  
We briefly discuss the second possibility in section 4.
Here, we take the first possibility.
We abandon the assumption of the gas
fraction being constant, and allow it to vary as a function of the
cluster mass, so that the observed luminosity--temperature relation is
reproduced (e.g., \cite{dsj93}). 
This is expected if non-gravitational processes affect hot gas
properties. 
The pre-heating model is an example (see e.g., \cite{bem01}),
and the variable galaxy formation efficiency is another example (see e.g.,
\cite{ptce00}; \cite{mtk01}).
If the galaxy formation efficiency is more
efficient in poor clusters than in rich systems, the gas fraction
increases with the cluster mass.
In order to reproduce the observed
luminosity--temperature relation, the gas fraction must be
proportional to $T_{\rm gas}^{1/2}$ ($M_{\rm cl}^{1/3}$):
\begin{equation}
 n_0 \propto M_{\rm cl}^{1/3}, ~~~
 r_{\rm c} \propto M_{\rm cl}^{1/3}, ~~~
 T_{\rm vir} \propto M_{\rm cl}^{2/3}.
\end{equation}
We use these relations in the following discussion.

Next, we consider the cooling effect.
We assumed the hot gas to be isothermal in the above discussion.
However, at the center, the cooling time may be shorter than the cluster
age and the cooling mechanism becomes important.
We define the cooling radius, $r_{\rm cool}$, as the radius where the
cooling time, $\tau_{\rm cool}$, equals the cluster
age, $\tau_{\rm age}$:
$\tau_{\rm cool}(r_{\rm cool}) = \tau_{\rm age}$. 
Here, the cooling time is defined as
\begin{equation}
 \tau_{\rm cool}(r) = \frac{3 n(r) k_B
  T_{\rm gas}}{\varepsilon(r)},
\end{equation}
where $\varepsilon = A n^2 T_{\rm gas}^{1/2}$ is the bolometric
emissivity and $A$ is a numerical constant.
Then, the cooling radius is described as
\begin{equation}
 r_{\rm cool} = r_c \left[ \frac{A n_0}{3 k_B T_{\rm gas}^{1/2}}
  \tau_{\rm age} - 1 \right]^{1/2}.
\end{equation}
It is to be noted that the value in the bracket does not depend on
the cluster mass, since $n_0 \propto T_{\rm gas}^{1/2}$.
Then, the gas number density at $r_{\rm cool}$ is given by
\begin{equation}
 n(r_{\rm cool}) = \frac{n_0}{[1+(r_{\rm cool}/r_c)^2]} =
  \frac{3 k_B T_{\rm gas}^{1/2}}{2 A \tau_{\rm age}}.
\end{equation}
Hereafter, $T_{\rm gas}$ denotes the gas temperature outside the
cooling radius. 
At $r < r_{\rm cool}$, the gas profile may deviate from the
$\beta$--model.
As a fiducial model, we take the self-similar gas density and
temperature profile within the cooling radius,
\begin{eqnarray}
 T(r) &=& T_{\rm gas} f_{\rm T} \left( \frac{r}{r_{\rm cool}}\right), ~~~
 n(r) = n(r_{\rm cool}) f_{\rm n} \left( \frac{r}{r_{\rm cool}}\right),
 \nonumber\\
 (&{\rm for}& ~~ r < r_{\rm cool}),
\end{eqnarray}
where $f_{\rm T}$ and $f_{\rm n}$ are functions independent of cluster
mass. We then obtain a relation between the luminosity within the
cooling radius, $L_{\rm cool}$, and the outer-region gas
temperature, $T_{\rm gas}$ (or the cluster mass  $M_{\rm cl}$), as 
\begin{eqnarray}
 L_{\rm cool} &=& \int_0^{r_{\rm cool}} 4 \pi r^2 \varepsilon (r) dr 
  \propto n(r_{\rm cool})^2 r_{\rm cool}^3 T_{\rm gas}^{1/2}
  \nonumber\\ 
 &\propto& T_{\rm gas}^3 \propto M_{\rm cl}^2.
\end{eqnarray}
That is, the ratio of $L_{\rm cool}$ to $L_{\rm bol}$ is constant.
In this case, the cut-off temperature, $T_{\rm cut}$, must be scaled to
$T_{\rm gas}$; then, it is expected that the ratio $T_{\rm
cut}/T_{\rm gas}$  is the same in all clusters.

At present, there is no compilation of the correlation between $L_{\rm
cool}$ defined above (or $L_{\rm cool}/L_{\rm bol}$) and $T_{\rm
gas}$. Therefore, in order to give a rough estimate, we use the data
compiled by Allen (2000, table 8) here. It is to be noted that
$L_{\rm cool, Allen}$ is estimated from the cooling flow model.
$L_{\rm cool, Allen}$ differs from $L_{\rm cool}$, defined above
in general, but we expect that there is a positive correlation between
$L_{\rm cool, Allen}$ and $L_{\rm cool}$.
His results showed that there is no clear trend between $L_{\rm cool,
Allen}/ L_{\rm bol}$ and $T_{\rm gas}$, although the scatter is
large.
This may indicate that clusters are roughly described by the scaling
relation [equation (6)].

\section{Heating Models}

The above discussion provides useful information for approaching the
``lack of the cold gas'' problem.
In this section, as an example, we consider the possibility that heating is
the main mechanism to solve the problem.
The heating model has one possible advantage in that it naturally
explains that there is no, or little, mass deposition at the cluster
center observationally.  Of course, there are problems;
for example, the required energy is huge.  If we assume a heating
source of $\Delta kT$ keV in the cluster core with a radius of
$r_{100} \times$ 100 kpc and a gas number density of
$n_3 \times 10^{-3}$ cm$^{-3}$ in $\tau_9$ Gyr, the required average
energy input rate is
\begin{eqnarray}
\frac{U}{\tau} &=& \frac{3n \Delta kT V}{\tau} \nonumber\\
&=& 1.9 \times 10^{43} \left(\frac{n_3}{10^{-3} {\rm cm^{-3}}}\right)
\left(\frac{\Delta kT}{1 \rm{keV}}\right) \nonumber\\
& & ~~~ \times \left(\frac{r_{100}}{100 \rm{kpc}}\right)^{3}
 \left(\frac{\tau_9}{1{\rm Gyr}}\right)^{-1} ~~{\rm erg~s^{-1}}.  
\end{eqnarray}
Since detailed modeling is beyond the scope of this Letter,
we do not discuss this problem any more.

Because cooling flow features are seen in many clusters, it is
reasonable to assume that they are in a steady state.  Within the
cooling radius, the cooling time is shorter than the cluster age, by
definition.  In order that clusters are in a steady state, the
cooling has to be balanced with heating. If clusters are
self-similar within the cooling radius [equation (6)], then the heating rate
has to be proportional to the gas temperature cubed (the cluster mass
squared) from the result obtained in the previous section.  
Hereafter, we assume that the heating rate, $\Gamma$, is described by
$\Gamma \propto M_{\rm cl}^p$ using a parameter, $p$.

When the exponent of $p$ equals to 2, clusters can become self-similar
within the cooling radius [equation (6)]. 
In this case, the ``lack of cold gas'' problem have to be observed in
all cooling flow clusters and the ratio
$T_{\rm cut}/T_{\rm gas}$ is the same in every cooling flow cluster. 

In the case of $p<2$, the heating affects the hot gas more
significantly in poor clusters than in rich clusters.  Thus, in
poor clusters, the ratio of the cut-off temperature to the gas
temperature outside the cooling radius becomes larger, that is,
the temperature decrement to the center becomes weaker. In rich
clusters, the simple isobaric cooling flow model may be a relatively good 
approximation.

On the contrary, in the case of $p>2$, the heating significantly
affects the hot gas in rich clusters.
Then, in poor clusters the ``lack of cold gas'' problem disappears
and the simple isobaric cooling flow model is applicable.

At present, several candidates for the heating source are proposed.
The above discussion can tell what kind of properties are required in the
heating sources.  The most popular candidate is supernova.
If we consider only normal star formation, the supernova rate is 
roughly proportional to the total cluster optical luminosity.
$L_{\rm cl}$, or the cluster mass: the heating rate due to 
supernova is proportional to the cluster mass, that is $p=1$.
In this case, the ``lack of cold gas'' problem should be observed
only in poor clusters.

Active Galactic Nuclei activity is another candidate.  The total
optical luminosity of a cD galaxy $L_{\rm cD}$ is proportional to
$L_{\rm cl}^{1.25}$ (\cite{bah00}).  If the activity of cD galaxies
(e.g., jet kinetic energy) is proportional to the optical
luminosity, $L_{\rm cD}$, then the heating rate follows $p \sim 1$.
There is another possibility.  The cD galaxies are more extended
than other giant elliptical galaxies; they show very extended
low surface brightness envelopes.  The luminosity of the cD
galaxy's envelope, $L_{\rm env}$, is proportional to
$L_{\rm cl}^{2.2}$ (\cite{bah00}).  If heating is
an accompanying result of the formation of the cD galaxy's envelope, then
the exponent becomes $p \sim 2$.  Then, clusters can become
self-similar within the cooling radius [equation (6)].

Next, we consider galaxy--galaxy collisions.
If the kinetic energy of galaxies is released in galaxy--galaxy
collisions, $\Gamma \sim N_{\rm gal} n_{\rm gal} \sigma_{\rm cl} (m_{\rm gal}
\sigma_{\rm cl}^2/2)$, where $N_{\rm gal}$ is the number of
galaxies in a cluster and $\sigma_{\rm cl}$ is the velocity
dispersion of the galaxies.
Since $N_{\rm gal} \propto M_{\rm cl}$ and $\sigma_{\rm cl}
\propto M_{\rm cl}^{1/3}$, the heating rate,
$\Gamma$, is $\sim M_{\rm cl}^2$; that is, $p=2$. 
In this case, clusters can become self-similar within the cooling
radius [equation (6)].

Finally, we consider heating by the magnetohydrodynamic effects proposed
by Makishima (see e.g., Makishima et al. 2001).
The galaxy motion causes significant magnetohydrodynamic
turbulence and frequent magnetic reconnection.
Since the kinetic energy of the galaxies is the origin of
the heating, the heating rate may be proportional to
$ N_{\rm gal} \sigma_{\rm cl}^2$. 
Then, the heating rate, $\Gamma$, is $\propto N_{\rm gal}
\sigma_{\rm cl}^2 / \tau_{\rm heating}$, where
$\tau_{\rm heating}$ is the heating time scale. 
Since we do not know the detail of the heating, we consider the following
two simple cases of the heating time scale.
One is that the heating time scale is proportional to the crossing time,
i.e., $R_{\rm cl}/\sigma_{\rm cl}$.
The other case is that the heating time scale is constant.
In both cases, the heating rate, $\Gamma$, is proportional to $M_{\rm
cl}^{5/3}$. 
Then, the ``lack of cold gas'' problem may be observed in most
clusters; the problem rather tends to become weak in poor clusters.

We only consider cooling flow clusters in the above discussion.
However, some candidates of the heating process (e.g., supernovae and
galaxy--galaxy collisions) also take place in the non-cooling flow
clusters. 
They may affect the hot gas properties (such as the temperature
structure) in the non-cooling flow clusters as well as in the
cooling flow clusters.
Observationally, there are small temperature inhomogeneities in
the non-cooling flow clusters, except for merging clusters.  This
indicates that the heating source does not exist in all clusters, but
exists only in the cooling-flow clusters.  That is, the heating
source may be closely related with the cooling properties.
If clusters are self-similar within the cooling radius [equation (6)],
the heating which accompanies the formation of cD galaxy's envelope is
promising, since the existence of cD galaxies is strongly correlated with
the existence of cooling flows.

\section{Discussion}

We considered the case that heating is an explanation of the
``lack of cold gas'' problem as an example.  Of course, there are
other possibilities.  If inhomogeneous metallicity is the
explanation, the cut-off temperature may be determined by the
degree of inhomogeneity in the metallicity, which may depend on
the history of each cluster.  Then, it is expected that the
strong $L_{\rm cool}$--$T_{\rm gas}$ relation does not show up
and the ratio $T_{\rm cut}/T_{\rm gas}$ changes from cluster to
cluster. We consider that observations from XMM-Newton can study
these features.  Next, we
consider a case in which the absorption by cold (neutral) material is the
explanation. In this case, the cut-off temperature is determined by
the amount of cold gas, which may be proportional to the accretion
rate, that is $L_{\rm cool}$. For brighter $L_{\rm cool}$ clusters,
the cut-off temperature may become higher.  We think this tendency
can be checked relatively easily in near-future observations.
It is a future
work to predict the relation between $L_{\rm cool}$ and $T_{\rm cut}$
for other cases.

In order to reproduce the observed \linebreak luminosity--temperature
relation, 
we made the gas fraction vary with the cluster mass in the above
discussion. 
There is another possibility that the he gas profile deviates from the
self-similar relation [equation (1)].
The self-similar relation [equation (1)] predicts that $r_{\rm c}
\propto R_{\rm cl} \propto T_{\rm gas}^{1/2}$, which is supported by
Vikhlinin et al. (1999) observationally;
but this point is still in controversy.
For example, Xu et al. (2001) showed that $r_{\rm c} \propto T_{\rm
gas}$ and Mohr and Evrard (1997) insisted that the X-ray isophotal size is proportional
to  $T_{\rm gas}$. 
These results indicate that non-gravitational processes are important
in cluster evolution (e.g., pre-heating model, \cite{eh91};
\cite{kai91}).  Deviations from the self-similar scaling relation
[equation (1)] give useful information to understand cluster physics.
If we consider these observed relations seriously,
our discussion must be changed. 
Therefore, we performed the same procedure as that in sections 2 and 3,
assuming $r_{\rm c} \propto R_{\rm cl} \propto T_{\rm gas}$ instead of
$n_0 = {\rm constant}$.  We obtained roughly the following relations:
\begin{equation}
 n_0 \propto M_{\rm cl}^{-1/2} \propto T_{\rm gas}^{-1}, ~~~
 T_{\rm gas} \propto M_{\rm cl}^{1/2}, ~~~
 L_{\rm bol} \propto T_{\rm gas}^{3/2}.
\end{equation}
Thus, in order to reproduce the observed \linebreak
luminosity--temperature 
relation we must abandon the assumption that the gas fraction is
constant.
The gas fraction must be proportional to $T_{\rm gas}^{3/4} \propto
M_{\rm cl}^{3/8}$, then, $n_0 \propto T_{\rm gas}^{-1/4}$.
We thus obtain $L_{\rm cool} \propto T_{\rm gas}^{27/8}$, or 
steeper, depending on the first term in the brackets of equation (4)
being much greater than 1 or not, instead of $L_{\rm cool} \propto
T_{\rm gas}^3$ in section 3.
If clusters are self-similar within the cooling radius [equation (6)],
the heating rate, $\Gamma$, has to be proportional to 
$M_{\rm cl}^{27/16}$, or steeper, instead of $M_{\rm cl}^2$ in section 3.
This difference may not significantly affect the rough discussion about
the heating source in this Letter.

On the other hand, recently, Ota (2001) suggested that clusters have
two different length scales of about 50 kpc and 200 kpc for each core
radius, assuming the Hubble constant to be $ 50 ~ {\rm km~s^{-1}
Mpc^{-1}}$.  Although we have not understood the origin of these two
length scales, here, we tentatively consider that the larger core
size describes the cluster properties.  Since the core size does
not correlate
with the temperature strongly, we assume $r_{\rm c}$ to be constant
as another
example (it is hard to consider cluster size is constant, thus we do
not assume $r_{\rm c} \propto R_{\rm cl}$) instead of $n_0 = {\rm constant}$.
Then, we obtain roughly
\begin{equation}
 n_0 \propto M_{\rm cl}^{2/3} \propto T_{\rm gas}, ~~~
 R_{\rm cl} \propto M_{\rm cl}^{1/3}, ~~~
 T_{\rm gas} \propto M_{\rm cl}^{2/3}.
\end{equation}
In this case, we obtain $L_{\rm bol} \propto T_{\rm gas}^{5/2}$,
thus the gas fraction must be proportional to $T_{\rm gas}^{1/4} \propto 
M_{\rm cl}^{1/6}$ in order to reproduce the observed
luminosity--temperature relation.
Then, $n_0 \propto T_{\rm gas}^{5/4}$ and we obtain $L_{\rm cool} \propto 
T_{\rm gas}^{21/8}$ or steeper depending on the first term in the
brackets of equation (4) being much greater than 1 or not.
If clusters are self-similar within the cooling radius [equation (6)], the
heating rate has to be proportional to $M_{\rm cl}^{7/4}$, or steeper.
This  difference may not significantly affect the rough discussion about 
the heating source, too. 

Recently, Chandra observations have shown that some clusters have a
complex X-ray surface brightness distribution, arising from the presence
of radio lobes (e.g., the Perseus cluster: \cite{fse00}; the Hydra A
cluster: \cite{mwn00}; A2052: \cite{bsm01}).
The size of the radio lobes is about several-times 10 kpc, which
corresponds to the size of the region where the ``lack of cold gas''
problem occurs.
At present, there is no evidence for shock-heated gas surrounding the
radio lobes. 
Therefore, the radio activity is not likely to be a major source of
energy input.
However, it is possible to affect our discussion through modifying the 
gas distribution. 
We think that the connection between the effect of radio lobes on the
surrounding gas and the ``lack of cold gas'' problem must be thought
about in future work. 

At present, the ``lack of cold gas'' problem has been reported in
only four clusters: A1795 ($T_{\rm vir} \sim 6 ~ {\rm keV}$,
\cite{tam01}), S\'{e}rsic
159-03 ($T_{\rm vir} \sim 3 ~ {\rm keV}$, \cite{kaa01}), A1835
($T_{\rm vir} \sim 8 ~ {\rm keV}$, \cite{pet01}) and the Virgo
cluster ($T_{\rm vir} \sim 2.5 ~ {\rm keV}$, \cite{mbfb01}).
The contradictory cases, i.e., cooling flow clusters without the
cut-off temperature, are not yet reported by XMM-Newton.
Thus, at present it may be premature to discuss the self-similarity; for
example, studying the relation between $T_{\rm cut}$ and $T_{\rm gas}$. 
However, it is worth pointing out that this problem has been reported
in a relatively low temperature cluster; S\'{e}rsic 159-03, and
the Virgo cluster.  Thus, this 
feature must be universal and this point needs to be taken into
account in considering the problem.  If heating is the explanation,
it is likely that the exponent of the heating, $p$, is around 2.

~~~

We would like to thank an anonymous referee for his/her useful comments.
We would also like to thank Prof. T. Ohashi for useful comments.
This research was supported by a Grant-in-Aid for Scientific
Research from the Ministry of Education, Science, Sports and Culture
(No. 12304009).

\end{document}